\documentclass[prl,aps,superscriptaddress,showpacs]{revtex4}

\usepackage{amssymb,amsmath}
\usepackage{graphicx}
\usepackage{nicefrac}

\begin{document}

\title
{Propagating mode-I fracture in amorphous materials using the continuous random network (CRN) model}
\author
{Shay I. Heizler}
\affiliation{Department of Physics, Bar-Ilan University,
 Ramat-Gan, IL52900 ISRAEL}
\affiliation{Department of Physics, Nuclear Research Center-Negev,
 P.O. Box 9001, Beer Sheva 84190, ISRAEL}
\author
{David A. Kessler}
\email{dave@ph.biu.ac.il}
\affiliation{Department of Physics, Bar-Ilan University,
 Ramat-Gan, IL52900 ISRAEL}
\author
{Herbert Levine}
\email{hlevine@ucsd.edu}
\affiliation{Department of Physics, University of California, San Diego, La Jolla, California 92093-0319}
 
\pacs{62.20.mm, 46.50.+a}

\begin{abstract}
We study propagating mode-I fracture in two dimensional amorphous materials using atomistic simulations.
We used the continuous random network (CRN) model of an amorphous material, creating  samples  using a
two dimensional analogue of the  WWW (Wooten, Winer \& Weaire) Monte-Carlo algorithm. For modeling fracture, molecular-dynamics 
simulations were run on the resulting
samples. The results of our simulations reproduce the main experimental features.
In addition to achieving a steady-state crack under a constant driving displacement (which had not yet been achieved by other atomistic models for amorphous materials), the runs show micro-branching, which increases with driving, transitioning to
 macro-branching for the largest drivings. Beside the qualitative visual similarity of the simulated
cracks to experiment, the simulation also succeeds in explaining the experimentally observed oscillations of the crack velocity.
\end{abstract}

\maketitle
 
The problem of dynamic fracture has been of great interest over the last two decades. Experiments, most of them
done in amorphous materials~\cite{fineberg_sharon0,fineberg_sharon1,fineberg_sharon2,fineberg_sharon3,fineberg_sharon4,review},
have shown the existence of steady-state propagating mode-I fracture, under an imposed constant driving strain.
The most interesting results concern the critical velocity at which a steady-state crack stops propagating on the
sample midline, and starts either to bifurcate or to create micro-branching. The onset of this behavior was found to occur at velocities dramatically smaller than those predicted by
Linear Elasticity Fracture Mechanics~\cite{freund,yoffe}.

Extensive theoretical efforts have been devoted to a variety of models in order to understand
the high velocity regime instability. In lattice materials,  steady-state cracks and the onset of instability
have been observed in simulations and the corresponding  solution for the  steady-state crack~\cite{slepyan,slepyan2} has been obtained,
 When the steady-state solution becomes
linearly unstable, the crack starts to oscillate or to bifurcate\cite{marderliu,pechenik,shay1,shay2}. However, the  behavior past the instability point did not
correspond closely to the experiments~\cite{fineberg_mar,shay1}.
This is presumably due to the mismatch between the anisotropic lattice employed in the models and the disordered isotropic (on long scales) nature of the experimental amorphous materials.
There has been some success in exploring the instability using phenomenological mesoscale approaches based on the phase-field~\cite{phase} or on cohesive zones~\cite{finite_element,gao_amor}; these however are difficult to quantitatively relate to an underlying microscopic picture.

Here, we focus instead on isotropic atomistic models. There have been a few earlier attempts in this direction~\cite{falk_langer,falk,tsviki,crn_eng}.
These have focused on one standard model for glass, the {\em binary-alloy model}; here, particles of two sizes with a 2-body (often Lenard-Jones) potential are mixed so as to avoid crystallization. Unfortunately,
these kind of models do not allow for freely propagating cracks at all, {\bf under a constant driving}. Only by continuously increasing the 
driving does the crack propagate, in direct contrast to experiment. In retrospect, this is not a surprising result, as this class of models yield large plastic zones.
Thus, these mixtures can be used for metallic glasses, but not for brittle materials.

Instead, we turn here to a different kind of model for describing amorphous materials, the
``continuous random network" (CRN). This model goes back to Zachariasen,~\cite{zacharainsen} who proposed it as a model for glass. In the CRN, the amorphous material is described as a ``corrupted" ordered material.
Wooten, Winer and Weaire~\cite{www,www2}  (WWW)  presented an  algorithm
for generating high-quality CRN's using Monte-Carlo techniques. Much effort has been expended in generating large 3D CRN's with thermodynamic properties matching those of several specific amorphous materials~\cite{vink}.
We note that Lu et. al.~\cite{crn_eng} has studied fracture in a small scale (1854 atoms) CRN model. However, they generated their network from a Poisson distribution of mass points, resulting in large voids, and their model did not exhibit steady-state fracture, but rather void coalescence.

We generated two-dimensional  CRN's by a 2D-analogue of the WWW algorithm (with $162\times272\approx44,000$ atoms in the largest sample).
The potential used in this constuction included both a 2-body central force and a 3-body bond-bending force:~\cite{potential}:
\begin{equation}
E_{\mathrm{tot}}=\sum_{i=1}^n\sum_{j=1\in nn}^3\left[\frac{1}{4}k_r(\vert \vec{r_{ij}}\vert-a)^2+
\frac{1}{2}k_{\theta}(\cos\theta_{ij}-\cos\theta_c)^2\right],
\label{potential}
\end{equation}
with $\theta_c=120^\circ$ and $a=4$. In our simulations, every  atom has exactly 3 neighbors.

\begin{figure}
\centering{
\includegraphics*[width=8cm]{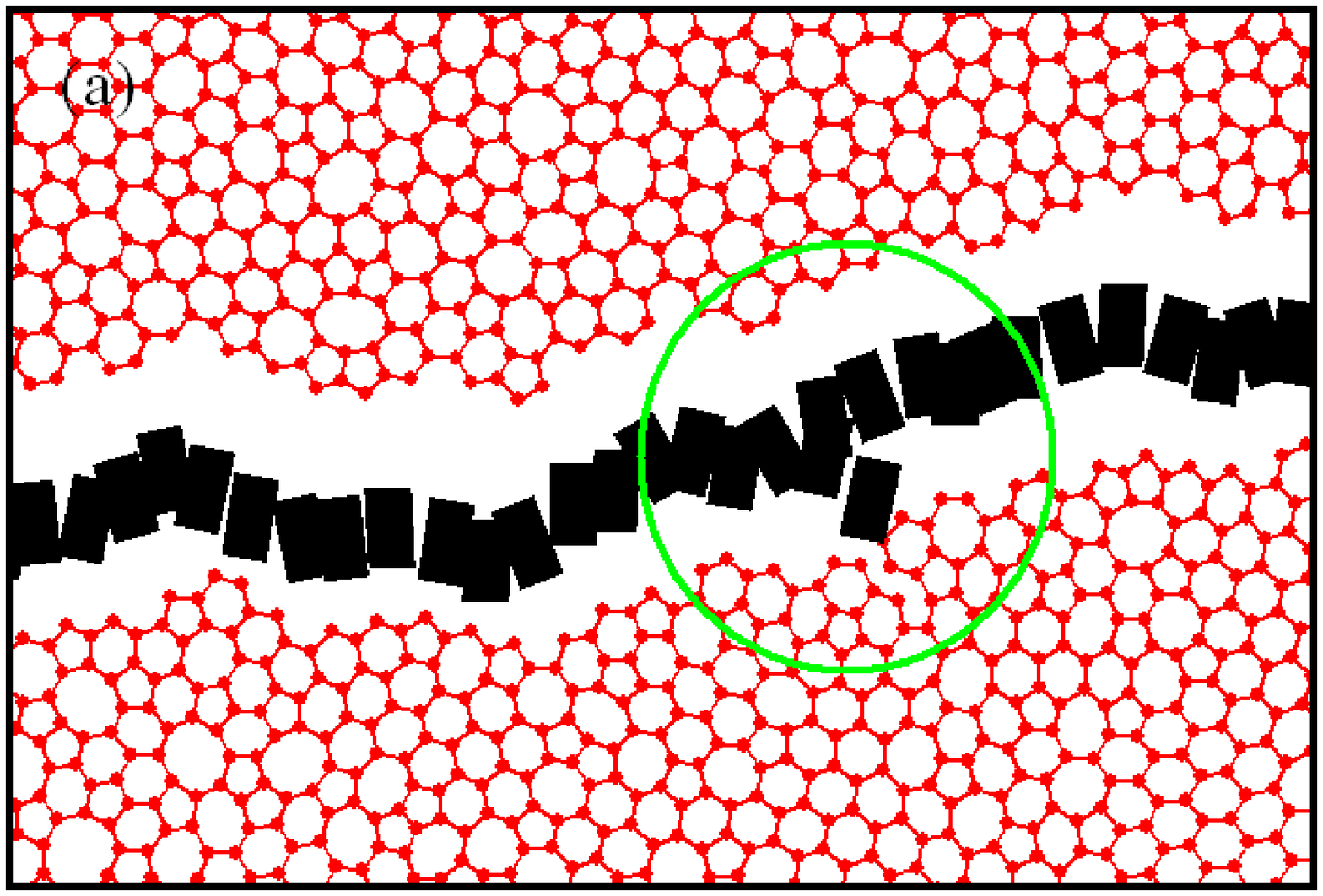}
\includegraphics*[width=8cm]{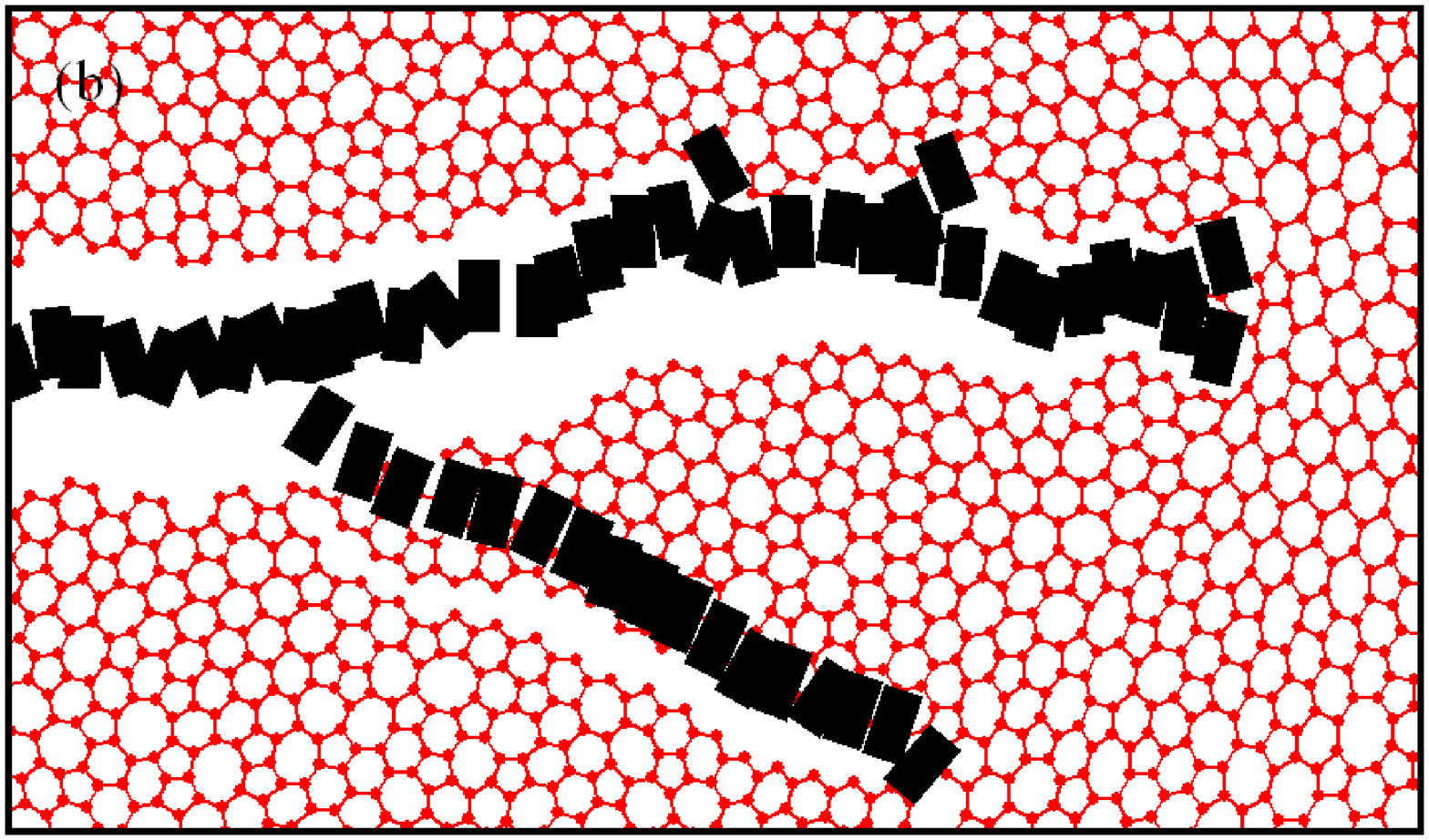}
}
\caption{(a) An image of a steady-state crack in the CRN model. We  see (in the circle) the origin of
micro-branching. The crack  reaches a ``tough zone"   and breaks parallel bonds, but
it does not have enough energy to create a visible micro-branch. $\eta=2$, $\nicefrac{\Delta}{\Delta_G}=2.7$
(b) At higher driving ($\eta=1.5$, $\nicefrac{\Delta}{\Delta_G}=3.1$), when the crack reaches the same tough zone, (displaced in this figure to enable showing its downstream development) the crack now has enough energy
to create two branches. Eventually one branch wins and continues as the main crack.
}
\label{fig4}
\end{figure}

Once we have constructed an appropriate amorphous material, we use an Euler scheme molecular-dynamics
simulation for the fracture study.
We use a piecewise-linear force law, in which the force decreases to zero immediately when the radial distance between
two atoms exceeds a certain threshold (which we set at $\varepsilon=1.25 a$). The variation of bond length in the material is sufficiently small that no bonds are broken in the uniformly stressed state, before the initial crack is seeded. In addition, we used a Kelvin-type viscosity
(with a viscose parameter $\eta$)~\cite{kess_lev2,kess_lev,shay1},
and the 3-body force law -- the gradient of Eq. \ref{potential}, that is removed when any of the involved bonds breaks.
These simulations are similar to those used in studies of lattice materials~\cite{marderliu,shay1,shay2}. We create
a constant strain in the CRN using a constant displacement ($\Delta$) and seed the system with an initial finite crack. Unlike
the binary-alloy model for glass, where the crack always arrests under constant driving~\cite{falk_langer,falk,tsviki},
in our CRN model, the crack, except for the smallest drivings,  always propagates through to the end of the sample, thereby
yielding a steady-state crack for small and moderate driving (Fig. \ref{fig4}(a)).  Only  close to the threshold Griffith's displacement, $\Delta_G$, does the crack occasionally fail to propagate.

When we increase the driving displacement, micro-branching starts to appear (Fig. \ref{fig4}(b)).
The simulation shown in Fig. \ref{fig4} demonstrates an important feature of fracture in amorphous materials. Unlike
lattice materials, %where there is a definite preferable direction for the crack to propagate (the midline of the sample),
in amorphous material% there is no such direction. 
the crack meanders off the midline.
When the crack reaches a ``tough zone", where there is no bond in the preferred direction for breaking, i.e., back to the midline, the crack starts
to bifurcate (Fig. \ref{fig4}(a) in the circle), but significant micro-branching will appear only if there is enough stored
energy in the mesh, i.e. Fig. \ref{fig4}(b). There does not appear to be a sharp threshold for micro-branching; rather, the extent of micro-branching increases continuously with the driving.

\begin{figure}[t]
\centering{
\includegraphics*[width=8cm]{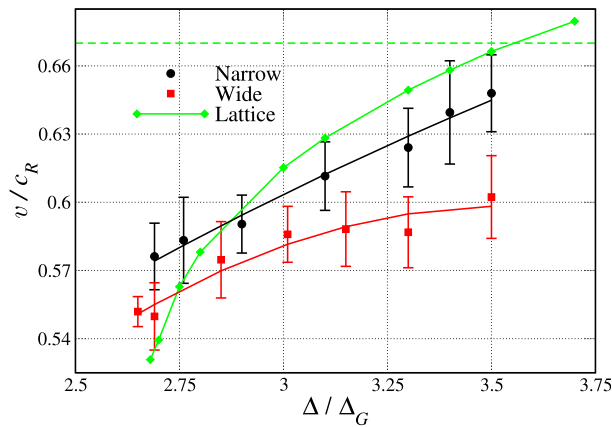}
}
\caption{(color online)The velocity of the crack, normalized to the Rayleigh velocity $c_R$, as a function of driving displacement,
$\Delta$, normalized to the critical Griffith displacement for crack propagation, $\Delta_G$, in two amorphous CRN's with
different widths (82 and 162 atoms in the $\hat{y}$-direction, correspondingly) and in a honeycomb lattice with the same force laws. The dashed line represents the critical velocity of the
lattice system. The amorphous systems have no well-defined threshold for micro-branching, and exhibit micro-branching
for $\Delta/\Delta_G$ well below the critical velocity of the lattice system. In all cases, $\eta=1.5$.
}
\label{v_del}
\end{figure}

It turns out  that the precise chosen crack path changes significantly with minor changes of our simulations, while the
macroscopic features (the $v(\Delta)$ curve, the length of the micro-branches, etc.) remains very much the same, albeit with some noise. We took advantage of this,
repeating the molecular dynamic simulations with the same driving conditions  several times (typically 20), changing the Euler time-step
each time by $3\%$, and averaging.  
Following the {\em main crack}, we measured the normalized velocity of the crack $\nicefrac{v}{c_R}$ as a function of the normalized
driving displacement $\nicefrac{\Delta}{\Delta_G}$. In Fig \ref{v_del}, we plot the $v(\Delta)$ curve for two (amorphous) CRN's with
different widths and for a lattice material. We can see that despite the fact that the two CRN's are completely different on the microscopic level, their
$v(\Delta)$ curves are quite close to each other. They are also of the same magnitude as the $v(\Delta)$ curve of the corresponding honeycomb lattice
material with the same force laws. In the lattice material, though, the critical velocity (the velocity when a steady-state crack stops propagating on the
midline of the sample and starts to create micro-branching) is approximately $0.67c_R$. In the CRN's there is no clearly defined threshold, 
and simulations exhibit micro-branching for $\Delta/\Delta_G$ well below the lattice critical velocity.
This result is directly analogous to the experimental finding that the critical velocity
of amorphous materials is much lower than that of ordered materials~\cite{review,cramer}.

\begin{figure}
\centering{
\includegraphics*[width=8.5cm]{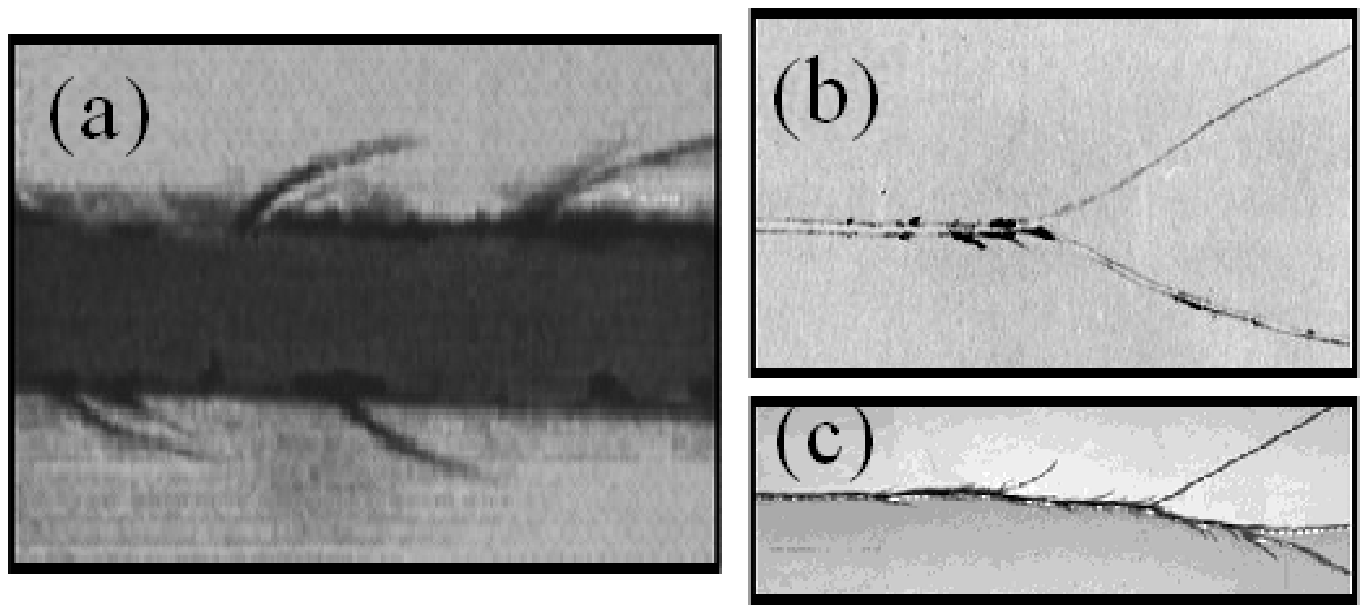}\\
\includegraphics*[width=8.5cm]{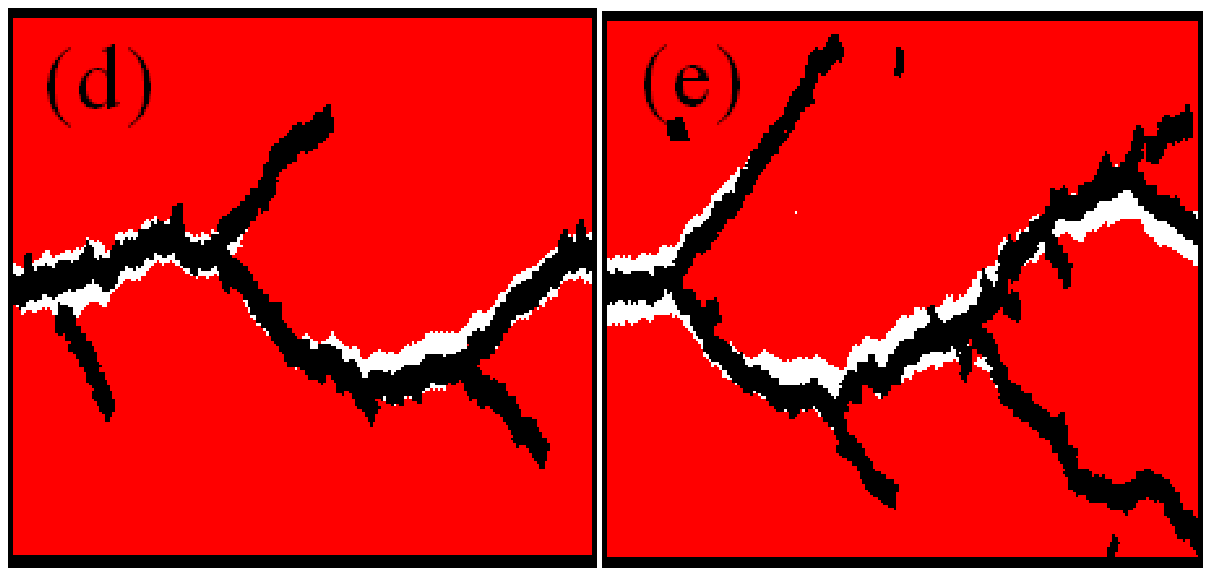}}
\caption{(color online)(a) An image of micro-branching in PMMA. The figure is taken from~\cite{fineberg_sharon2}. (b)-(c) Images
of macro-branching in experiments. Image (b) is in Homalite-100 (glassy polymer) and taken from~\cite{macro_nisuy}
and (c) is in PMMA and taken from~\cite{fineberg_sharon4}. (d) An image of micro-branching fracture in our CRN simulation.
$\eta=1.5$, $\nicefrac{\Delta}{\Delta_G}=3.1$.
(e) In higher driving ($\eta=1.5$, $\nicefrac{\Delta}{\Delta_G}=4$),
macro-branching begins to appear in our CRN simulation; both branches of the crack reach the end  of the sample.
}
\label{fig6}
\end{figure}

As seen in Fig. 3, the CRN's behavior beyond the point of instability compares well 
with that of a propagating crack in experiments on a mode-I fracture
in amorphous materials~\cite{fineberg_sharon2,fineberg_sharon4,macro_nisuy}, as seen in Fig. \ref{fig6}.
Fig. \ref{fig6}(a) shows a photo of an experiment in PMMA. Micro-branching is seen clearly in addition to the
main crack. Fig. \ref{fig6}(d) shows a parallel image from our simulation (of course, at a different length scale;
our simulation is atomistic). Qualitatively, the images look quite similar. In the experiments, increasing the driving
causes a bifurcation of the main crack to several (at least two) macro-branches traveling to the end of the sample,
both in Homalite-100 (Fig. \ref{fig6}(b)), and in PMMA (Fig. \ref{fig6}(c)). Exactly the same feature is seen in
our simulations (Fig. \ref{fig6}(e)).

\begin{figure}
\centering{
\includegraphics*[width=7cm]{v_t}
\includegraphics*[width=7cm]{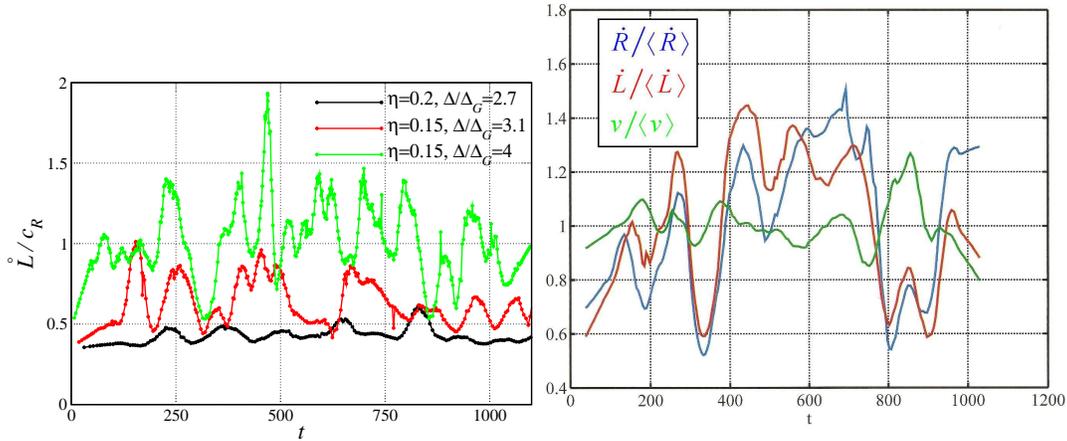}
}
\caption{(color online)(a)The rate of change of fracture surface, $\dot{L}$, normalized to the Rayleigh wave speed, $c_R$, as a function of the time for different drivings. For small
driving, the curve is quite flat. At larger driving the curve starts to oscillate, due to the appearance of micro or macro-branching.   (b) Correlation test
to the crack's velocity oscillations. The oscillations of the electrical resistance behave like
$\dot{L}(t)$, and not like the main crack velocity.  Here $\eta=1.5$, $\Delta/\Delta_G=3.5$.  For both panels, the width is 162 atoms.
}
\label{fig7}
\end{figure}

One of the interesting features revealed in the experiments is the purported increase in the amplitude of oscillations of the crack velocity with increasing  driving (or increasing average velocity). In the steady-state
regime, the $v(t)$ curve is quite smooth. When the driving increases and micro-branching starts to appear, oscillations begin to appear in the electrical resistance of a conductive layer; attached to the sample~\cite{fineberg_sharon2,fineberg_sharon4,review}. This was then interpreted as an oscillation of the crack velocity, based on a calibration of the resistance measurement, either by simulation of the conductive
layer~\cite{marder_jay1,marder_jay2,Ravi_Chandar},
or experimentally by quasi-static crack measurements of a single straight crack~\cite{fineberg_sharon2}. 
Our simulations do not show an increase in oscillations of the velocity of the {\em main crack} as the crack velocity increases. Instead, we found increasing oscillations in the rate of change of the overall length of the fracture surface, $\dot{L}(t)$
where $L(t)\equiv \varepsilon\sum_{t_0}^{t_i}n_i(t)$ and $n_i$ is one for a cracked bond. In Fig. \ref{fig7}(a) we present three time histories of this new velociity. The black curve is for an essentially steady-state crack,
the red curve corresponds to a crack with a small degree of micro-branching  (Fig. \ref{fig6}(d)) and the green curve is for a case when macro-branches were
created (Fig. \ref{fig6}(e)). 
We can clearly see that the steady-state curve is much smoother than the other
curves. In addition, the shape of the oscillations in the velocity curves are very similar to what was
seen in the experiments~\cite{fineberg_sharon2,fineberg_sharon4,review}.

\begin{figure}
\centering{
\includegraphics*[width=8cm]{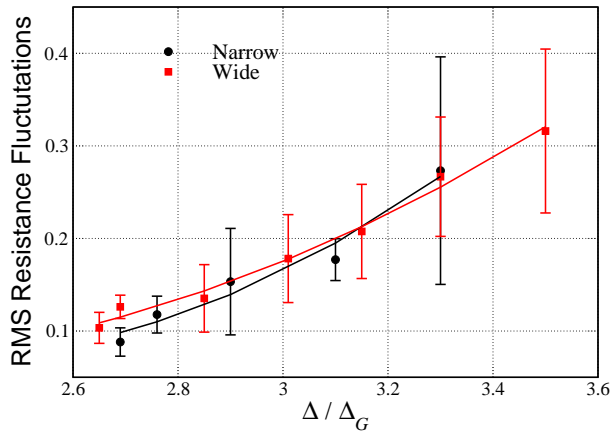}
}
\caption{(color online)The amplitude of the RMS fluctuations in the oscillations of the  electrical resistance during the course of the simulation as a function of the normailized driving $\Delta/\Delta_G$.  The ``wide" system is 162 atoms wide, and the narrow one 82 atoms wide.  In both cases, $\eta=1.5$.
}
\label{rrms}
\end{figure}

It is straightforward to model the conductive layer by a grid of identical resistors, with the resistors removed whenever they cross a broken bond.  Then for a specific crack configuration in our CRN we solve a Laplace-like equation for the voltage in the system, 
from which we can calculate the current density,
and determine the sample's electrical resistance. 
A  calculation of the  resistance in our cracked CRN shows that it
behaves like $\dot{L}(t)$, and not like the main crack's velocity (Fig. \ref{fig7}(b)).
Plotting the RMS amplitude of the oscillations of the electrical resistance (Fig. \ref{rrms}) shows significant growth of the oscillations 
with the driving displacement, as in the experiments~\cite{fineberg_sharon2}.

In conclusion, the main features of the micro-branching instability at large driving observed
in the experiments were reproduced in our CRN-grid simulations both qualitatively (Fig. \ref{fig6}) and
semi-quantitatively (Fig. \ref{rrms}). We should note that a fundamental question that our simulations do not resolve is the role of the third dimension
in the experiments. The experiments on gels~\cite{gels},
show systematic changes with the thickness of the gel, indicating the possible importance of such effects.
We hope to address this issue after extending our model to three dimensions, which should in principle be straightforward,
although computationally demanding. Also in the future, we plan to investigate  the scaling properties of our CRN cracks as the system size is scaled up.

%\begin{acknowledgments}
%\end{acknowledgments}


\begin{thebibliography}{99}
\bibitem{fineberg_sharon0}E. Sharon, S. P. Gross and J. Fineberg, {\em Phys. Rev. Lett.} {\bf 74}, 5096 (1995).
\bibitem{fineberg_sharon1}E. Sharon, S. P. Gross and J. Fineberg, {\em Phys. Rev. Lett.} {\bf 76}, 2117 (1996).
\bibitem{fineberg_sharon2}E. Sharon and J. Fineberg, {\em \prb} {\bf 54}, 7128 (1996)
\bibitem{fineberg_sharon3}E. Sharon and J. Fineberg, {\em Philos. Mag. B} {\bf 78}, 243 (1998)
\bibitem{fineberg_sharon4}E. Sharon and J. Fineberg, {\em Advanced Engineering Materials}, {\bf 1}, 119 (1999)
\bibitem{review}J. Fineberg and M. Marder, {\em Phys. Repts.} {\bf 313}, 2 (1999).
\bibitem{freund}L.B. Freund, {\em Dynamic Fracture Mechanics}, Cambridge
University Press (1998).
\bibitem{yoffe}E.H. Yoffe, {\em Philos. Mag.} {\bf 42}, 739 (1951).
\bibitem{slepyan}L. I. Slepyan, {\em Doklady Akademii Nauk SSSR}, {\bf 258},
561 (1981) $[${\em Sov. Phys. Dokl.} {\bf 26}, 538 (1981)$]$.
\bibitem{slepyan2}Sh. A. Kulamekhtova, V. A. Saraikin and L. I. Slepyan,
{\em Mech. Solids} {\bf 19}, 102 (1984).
\bibitem{marderliu}M. Marder and X. Liu, {\em \prl} {\bf 71},
2417 (1993).
\bibitem{pechenik}L. Pechenik, H. Levine and D. A. Kessler,
{\em J. Mech.  Phys. Solids} {\bf 50}, 583 (2002).
\bibitem{shay1}S.I. Heizler, D.A. Kessler and H. Levine, {\em \pre}, {\bf 66}, 016126 (2002).
\bibitem{shay2}S.I. Heizler and D.A. Kessler, {\em Contin. Mech. Thermodyn.} {\bf 22}, 505  (2010).
\bibitem{fineberg_mar} M. Marder and J. Fineberg, {\em Phys. Today} {\bf 49}, 24 (1996).
\bibitem{phase}H. Henry, {\em EuroPhys. Lett.} {\bf 83}, 16004 (2008).
\bibitem{finite_element}C. Linder and F. Armero, {\em Finite Elements in Analysis and Design}, {\bf 45}, 280 (2009).
\bibitem{gao_amor}H. Gao and P. Klein, {\em J. Mech. Phys. Solids} {\bf 46}, 187 (1998).
\bibitem{falk_langer}M.L. Falk and J.S. Langer, {\em \pre} {\bf 57}, 7192 (1998).
\bibitem{falk}M.L. Falk, {\em \prb} {\bf 60}, 7062 (1999).
\bibitem{tsviki}T.Y. Hirsh and D.A. Kessler, arXiv:cond-mat/0409607, (2004).
\bibitem{crn_eng}Q. Lu, N. Marks, G.C. Schatz and T. Belytschko, {\em \prb}, {\bf 77}, 014109 (2008).
\bibitem{zacharainsen}W.H. Zachariasen, {\em J. Am. Chem. Soc.} {\bf 54}, 3841 (1932).
\bibitem{www}F. Wooten, K. Winer and D. Weaire, {\em \prl} {\bf 54}, 1392 (1985).
\bibitem{www2}F. Wooten and D. Weaire, {\em Sol. State Phys.}, {\bf 40}, 1 (1987).
\bibitem{vink}R.L.C. Vink, {\em Computer Simulations of Amorphous Semiconductors},
Ph.D. Thesis, Utrecht University (2002).
\bibitem{potential}Y. Tu, J. Tersoff, G. Grinstein and D. Vanderbilt, {\em \prl} {\bf 81}, 4899 (1998). 
\bibitem{kess_lev2}D.A. Kessler and H. Levine, {\em \pre} {\bf 59}, 5154 (1999).
\bibitem{kess_lev}D.A. Kessler and H. Levine, {\em \pre} {\bf 63}, 016118 (2000).
\bibitem{cramer}T. Cramer, A. Wanner and P. Gumbsch, {\em Phys. Rev. Lett.} {\bf 85}, 788 (2000).
\bibitem{macro_nisuy}M. Ramulu and A.S. Kobayashi, {\em Int. J. Fract. Mech.} {\bf 27}, 187 (1985).
\bibitem{marder_jay1}J. Fineberg, S. P. Gross, M. Marder and H. L. Swinney, {\em Phys. Rev. Lett.} {\bf 67}, 457 (1991).
\bibitem{marder_jay2}J. Fineberg, S. P. Gross, M. Marder and H. L. Swinney, {\em Phys. Rev. B}, {\bf 45}, 5146 (1992).
\bibitem{Ravi_Chandar}D. Bonamy and K. Ravi-Chandar, {\em Int. J. Frac.} {\bf 134}, 1 (2005).
\bibitem{gels}A. Livne, O. Ben-David and J. Fineberg, {\em \prl} {\bf 98}, 124301 (2007).
\end{thebibliography}
\end{document}